\documentclass[11pt]{article}
\usepackage{amsfonts,amssymb,epsfig,amsmath}
\usepackage{graphicx}
\usepackage{color}
\renewcommand{\text}[1]{#1}

\newcommand{\be}{\begin{equation}}
\newcommand{\ee}{\end{equation}}
\newcommand{\ben}{\begin{displaymath}}
\newcommand{\een}{\end{displaymath}}
\newcommand{\bea}{\begin{eqnarray}}
\newcommand{\eea}{\end{eqnarray}}
\newcommand{\bean}{\begin{eqnarray*}}
\newcommand{\eean}{\end{eqnarray*}}
\newcommand{\nn}{\nonumber \\}
\newcommand{\ba}{\begin{array}}
\newcommand{\ea}{\end{array}}
\newcommand{\bi}{\begin{itemize}}
\newcommand{\ei}{\end{itemize}}

\renewcommand{\theequation}{\arabic{section}.\arabic{equation}}
\def\theequation{\thesection.\arabic{equation}}

\def\g{\gamma}
\def\G{\Gamma}

\def\G{\Gamma}
\def\g{\gamma}

\def\s{\sigma}

\newcommand{\ft}[2]{{\textstyle\frac{#1}{#2}}}

\begin{document}

\makeatletter
\renewcommand{\theequation}{\thesection.\arabic{equation}}
\@addtoreset{equation}{section}
\makeatother

\begin{titlepage}

\vfill
\begin{flushright}
KIAS-P10040
\end{flushright}

\vfill

\begin{center}
   \baselineskip=16pt
   {\Large\bf $3D$ gauged supergravity from wrapped M5-branes with\\[1ex] AdS/CMT applications. 
   }
   \vskip 2cm
     Eoin \'O Colg\'ain$^{1}$ \& Henning Samtleben$^{2}$
       \vskip .6cm
             \begin{small}
      \textit{$^{1}$ Korea Institute for Advanced Study, \\
        Seoul 130-722, Korea \\  \vspace{3mm}
$^{2}$ Universit\'e de Lyon, Laboratoire de Physique, UMR 5672, CNRS,\\Ecole Normale Sup\'erieure de Lyon,
F-69364 Lyon cedex 07, France,\\
Institut Universitaire de France}
        \end{small}\\*[.6cm]
\end{center}

\vfill
\begin{center}
\textbf{Abstract}\end{center}

\begin{quote}
 By identifying a bosonic consistent truncation from the $\tfrac{1}{4}$-BPS wrapped M5-brane geometry of Maldacena, Strominger and Witten in $D=11$ supergravity and finding a supersymmetric extension, we recover an $N=2$ $D=3$ supergravity theory. Reductions of a large class of supersymmetric solutions corresponding to wrapped M2 and M5-branes lead to black strings and warped $AdS_3$ solutions preserving supersymmetry. With a view to AdS/CMT applications, we also construct a numerical hairy BTZ black hole and, as a preliminary step in this direction, determine the conductivity of the dual CFT.
\end{quote}
\vfill

\end{titlepage}

\section{Introduction}
Continued interest in the AdS/CFT correspondence owes much to the potential computational control it gives in studying strongly coupled, non-perturbative field theories. Having witnessed attempts to model QCD-like theories holographically, much focus has shifted to condensed matter theory (CMT). In contrast to its predecessor, AdS/QCD, swift progress in AdS/CMT has been precipitated through simple, bottom-up models with a phenomenological flavour. With attention now turning to the task of embedding promising bottom-up theories in higher-dimensional supergravity, a footing where we have greatest confidence in the AdS/CFT, we have seen a small renaissance in the study of consistent truncations.

An interesting spin-off of all this recent activity is a deeper understanding of how the various supergravity theories are related. In particular, starting from simplified setting of $D=11$ supergravity, it would be attractive to map out all the lower-dimensional supergravities arising through consistent truncations. Prior to AdS/CMT research, the most well known examples in either Type IIB or $D=11$ involved sphere reductions to maximally supersymmetric theories on $S^5$ \cite{s51,s54}, $S^7$ \cite{dewit} and  $S^4$ \cite{s41}. A few years ago it was shown that the most general supersymmetric $AdS_5$ solutions of $D=11$ supergravity \cite{ads5M6} (of which Sasaki-Einstein $Y^{p,q}$ \cite{ypq} belong) permits a consistent reduction to minimal ${N}=2$ $D=5$ gauged supergravity \cite{d5min1,d5min2}. Since then there have been subsequent studies on consistent reductions \cite{jerome1,oscar}, while the extension to examples incorporating massive modes appeared in \cite{MMT} and \cite{se7mass}.

Then since the explosion in interest in AdS/CMT, there has been a blistering race to embed promising bottom-up, holographic superconducting models \cite{holsc1,sean2} in Type IIB \cite{embedIIB} and $D=11$ \cite{embedM,cmjerome2}. The interest in $SE_5$ bosonic reductions from Type IIB peaked earlier this year when numerous overlapping results appeared \cite{1se5} (see also \cite{Liu1,T11}) . Shortly after the overlooked fermion reduction on Sasaki-Einstein spaces from Type IIB \cite{Bah10,Liu2} and $D=11$ \cite{Bah11} also appeared in the literature.

In terms of reductions to $D=3$, the story is not so well established. Reductions from $N =1$ $D=6$ supergravity \cite{Lu1} aside, in this paper we present one of the first examples from $D=11$\footnote{A consistent truncation covered by our ansatz focusing on supersymmetric G\"odel spacetimes appeared in \cite{Levi}.}. We will begin with the geometry dual to chiral $\mathcal{N} = (4,0)$ SCFT in two-dimensions. The dual spacetime, commonly referred to by the Maldacena, Strominger, Witten (MSW) geometry, arises from M5-branes wrapping K\"{a}hler four-cycles in a Calabi-Yau three-fold $CY_3$  \cite{MSW}:
\bea
\label{MSW}
ds^{2} &=& \frac{1}{m^2} ds^{2}(AdS_3) + \frac{1}{4 m^2} ds^2(S^2) + ds^2(CY_3), \nn
F^{(4)} &=& \pm \frac{1}{2 m} vol(S^2) \wedge J.
\eea
Here $m$ denotes the inverse $AdS_{3}$ radius, $J$ the K\"{a}hler form and the choice of sign depends on whether one wraps M5-branes or anti-M5-branes. This geometry with $SU(3)$-holonomy appeared as the only example in a recent classification of $AdS_3 \times S^2$ geometries in $D=11$ with $SU(3)$-structure and eight supercharges \cite{eoin1}. Given its uniqueness, the strategy then is to deform this geometry in a similar way to \cite{M5adscmt} where, in comparison to the case presented here, the background is sourced by wrapped M5-branes. Next one imposes the $D=11$ equations of motion to derive a set of $D=3$ equations of motion and then reconstructs the Lagrangian. Such a process leads to a consistent truncation on $S^2 \times CY_3$ and a scalar potential that admits a single supersymmetric vacuum inherited from the parent MSW geometry. For simplicity and also with one eye on the AdS/CMT literature where $U(1)$ isometries are ubiquitous, we have chosen not to gauge the $S^2$ in the reduction, though one imagines that this can be incorporated.

As the general structure of three-dimensional supergravity theories with various amounts of supersymmetry
has been explored in detail in the past \cite{Nicolai:2000sc,deWit:2003ja},
the form of the bosonic Lagrangian that we obtain in the reduction
allows to identify the underlying $N=2$ structure. After redualizing the three-dimensional vector 
degrees of freedom, the field content of the three-dimensional theory is given by three complex scalar multiplets, parametrizing the K\"ahler manifold $SU(2,1)/U(2)\times SU(1,1)/U(1)$. 
Supersymmetric interactions are triggered by a particular shift symmetry gauging and a holomorphic superpotential. This analysis allows to reconstruct the full fermionic sector, including the Yukawa couplings and the Killing spinor equations without performing the explicit fermionic reduction. In particular, the Killing spinor equations may be employed to construct supersymmetric solutions directly in three dimensions.

Having performed the reduction and identified a supersymmetric extension, in the rest of the paper, we reduce a large class of supersymmetric wrapped M2 and M5-brane \cite{braneBH,3chargeBH} geometries to $d=3$ in order to check the consistency. We observe that the near-horizon limit of one of these classes recovers a null-warped $AdS_3$ spacetime in the same context as \cite{sinha1}, while black string solutions reminiscent of \cite{d3bstring} also appear. We also construct numerically a black hole solution that is asymptotically the same as charged BTZ \cite{BTZ,chargedBTZ} with the $CY_3$ breathing mode providing the hair. Using this background, as a preliminary foray into the AdS/CMT world, we determine the conductivity of the dual system recovering results similar to \cite{Maity,Hung}, but stress our black hole has an M-theory embedding, even away from the probe limit. 

A big future challenge for this geometry concerns whether it captures signatures of the Luttinger liquid model \cite{Luttliq}. In contrast to higher dimensions, Fermi-Landau liquid theory is not applicable in one spatial dimension, and for theories with a gapless branch, the effective description of the physics may be given in terms of a Luttinger liquid. One of the most interesting predictions of this model is an experimentally observed \cite{scsexp} effect known as \textit{spin-charge separation} \cite{spincharge}, where the spin and charge excitations of spin-half fermions travel at different velocities.  To date, we have already seen attempts to model Luttinger liquids holographically \cite{Maity,Hung}. However, neither group were able to report any signature of spin-charge separation, an effect that would be an important milestone to reproduce from a gravity set-up if one is going to consider AdS/CMT more seriously in lower-dimensions.

\section{Consistent Truncation}
In this section we give an account of the consistent truncation forming the bedrock of this study. We will consider an Abelian reduction of the MSW \cite{MSW} geometry using the simple ansatz
\bea
\label{red_ansatz}
ds^2 &=&  ds^{2}(M_3) + e^{2 U} ds^2(S^2) + e^{2 V} ds^2(CY_3), \nn
F^{(4)} &=& \alpha\, vol(S^2) \wedge J + vol(S^2) \wedge H_1 +  \beta J^2 + J \wedge H_2  \nn  &+& \gamma\, vol(M_3) \wedge \cos \theta d \phi + \left[ d f \wedge \Omega + c.c.\right],
\eea
where $\theta, \phi$ parameterise the two-sphere and $J, \Omega$ denote the usual forms on $CY_3$. The scalars $\alpha, \beta, \gamma, U, V$ are \textit{a priori} real scalars, $f$ is a complex scalar, and $H_1, H_2$ define two-forms on $M_3$. As we shall see shortly, only $U, V, f$ survive the reduction with $\alpha$ being demoted to a constant in the process.
 
Note also that the original solution is recoverable from this ansatz when $M_3$ is an $AdS_3$ metric of unit inverse radius $m=1$, $ \alpha = \tfrac{1}{2}$ and $ U = - \ln 2$. The immediate task now is to solve the eleven-dimensional equations of motion
\bea
d F^{(4)} &=& 0, \\
d (* F^{(4)} ) &=&  - \tfrac{1}{2} F^{(4)} \wedge F^{(4)}, \\
R_{MN} &=& \tfrac{1}{12} F^{(4)}_{MP QR} F_{N}^{(4)~PQR} - \tfrac{1}{144} g_{MN} F^{(4)}_{PQRS} F^{(4)~PQRS},
\eea
where $M,N = 0, \cdots 10$. While the details of the reduction may be found in appendix A, the $D=3$ equations of motion that one obtains may be derived from the following Lagrangian
\bea
\mathcal{L}_3 &=& e^{2 U + 6 V} \left[R vol_3 + 2 d U \wedge * dU + 24 d U \wedge * d V + 30 d V \wedge * dV  \right] \nn &-& 8 e^{2 U} df \wedge * df^* - \tfrac{1}{2} e^{-2U + 6V} H_1 \wedge * H_1 - \tfrac{3}{2} e^{2 U + 2 V} H_2 \wedge * H_2  \nn
&-&   8 i B_1 \wedge df \wedge df^* - 3 \alpha B_2 \wedge H_2 + (2 e^{6V} - \tfrac{3}{2} \alpha^2 e^{-2 U + 2V}) vol_{3},
\eea
where we have introduced the gauge potentials $B_1, B_2$ such that $H_1 = d B_1, H_2 = d B_2$.

One may also find the Einstein frame Lagrangian by performing the metric rescaling $ g_{\mu \nu} = e^{-2(2U+6V)} \hat{g}_{\mu \nu}$. The Einstein-frame Lagrangian is \bea \label{einsteinL}
\mathcal{L}_{\textrm{Einstein}} &=& \hat{R} \hat{vol}_3 - 6 d U \wedge \hat{*} dU - 24 d U \wedge \hat{*} d V -42  d V \wedge \hat{*} dV  \nn &-& 8 e^{- 6 V} df \wedge \hat{*} df^* - \tfrac{1}{2} e^{12 V} H_1 \wedge \hat{*} H_1 - \tfrac{3}{2} e^{4 U + 8 V} H_2 \wedge \hat{*} H_2 \\
&-& 8 i B_1 \wedge df \wedge df^*  - 3 \alpha B_2 \wedge H_2 + (2 e^{-6 U - 12 V } - \tfrac{3}{2} \alpha^2 e^{-8 U -16 V}) \hat{vol}_{3}.  \nonumber
\eea

One may then proceed to simplify the action by diagonalising the scalar kinetic terms
\be
W =  U + 2 V,
\ee
resulting in
 \bea \label{diageinsteinL}
\mathcal{L}_{\textrm{Einstein}} &=& \hat{R} \hat{vol}_3 - 6 d W \wedge \hat{*} d W - 18  d V \wedge \hat{*} dV  \nn &-& 8 e^{- 6 V} df \wedge \hat{*} df^* - \tfrac{1}{2} e^{12 V} H_1 \wedge \hat{*} H_1 - \tfrac{3}{2} e^{4 W} H_2 \wedge \hat{*} H_2 \\
&-& 8 i B_1 \wedge df \wedge df^*  - 3 \alpha B_2 \wedge H_2 + (2 e^{-6 W } -  \tfrac{3}{2} \alpha^ 2 e^{-8 W}) \hat{vol}_{3}.  \nonumber
\eea

Having established the form of the Einstein frame Lagrangian, one may determine the vacua of the theory by analysing the critical points of the scalar potential
\be
\mathcal{L}_{\textrm{pot}} = (2 e^{-6 W } - \tfrac{3}{2} \alpha^2 e^{-8 W}) \hat{vol}_{3}.
\ee
One finds that a single $AdS_3$ vacuum exists when $W = \ln (\alpha) $. This is no surprise, as setting $V=0$, $\alpha = \tfrac{1}{2}$, we recover the supersymmetric vacuum of the original MSW solution (unit $AdS_3$ radius). This provides us with an initial non-trivial consistency check.

As an extra consistency check, we can start from the consistent reduction from eleven dimensions to five dimensions on $KE_6$ appearing in \cite{KE6}\footnote{ Although the stability of non-supersymmetric solutions is a constant concern, when $KE^6 \equiv \mathbb{CP}^3$, the $AdS_5$ vacuum appears to be classically stable \cite{KE6stability}. } by truncating out the complex scalar $f$ i.e. $f=0$. We recall that the Ricci tensor for the $KE_6$ metric is simply
\be
R_{mn} = 2 c^2 g_{mn},
\ee
so $c=0$ makes the $KE_6$ space $CY_3$. To make connection with the work appearing above, one simply has to further reduce the five-dimensional theory to three-dimensions. This may be done by employing the identifications
\bea
\label{KE6flux}
\tilde{H}_{4} = vol(S^2) \wedge H_1, \quad \tilde{H}_{2} = \alpha\, vol(S^2) + H_2,
\eea
where we have introduced tilde notation to differentiate. Then by relabeling the warp factor $U$ of \cite{KE6} as $V$, one recovers the above equations of motion. To be a little more precise, (2.5), (2.6) and (2.7) of \cite{KE6} are respectively, (\ref{fluxeq1}), (\ref{fluxeq2}) and (\ref{diffeq2}) of this draft. Then, one can derive (\ref{eqn3}) from the Einstein equation (2.8) of \cite{KE6}, while (\ref{diffeq1}) can also be extracted from the sphere directions of the Einstein equation by using the Christoffel symbols  $\G^{\mu}_{\theta \theta} = \G^{\mu}_{\phi \phi} = - \partial_{\rho} U \partial^{\rho} V$.

\section{Effective $D=3$ supergravity}

In this section we will cast the three-dimensional action obtained in the reduction into the standard form of
three-dimensional gauged supergravity. In particular, this allows to reconstruct the entire fermionic sector as well as
the Killing spinor equations from the underlying $N=2$ supersymmetry structure.

\subsection{Redualizing and scalar geometry}

In order to make the underlying $N=2$ structure of (\ref{diageinsteinL}) manifest, we first need to dualize all propagating
degrees of freedom into the scalar sector~\cite{deWit:2003ja}. The bosonic equations of motion for the vector fields
\bea
0 &=& d (e^{12 V} * H_1) + 8 i df \wedge d {f}^*\;,  \nonumber\\
0 &=& d(e^{ 4 W} * H_2) +2 \alpha H_2\;,
\label{eomH1Vf}
\eea
allow to introduce scalar fields $X, Y$ according to the duality equations
\bea
H_1 &=& e^{-12V}\,\left\{ *dX+4i(f *df^* - f^* *df) \right\}
\;, \nonumber\\
H_{2} &=& e^{-4 W } *DY
\;,
\label{defH12}
\eea
with the covariant derivative defined as $D_\mu Y = \partial_\mu Y - 2\alpha B_2{}_\mu$\,.
The integrability conditions of (\ref{defH12}) reproduce (\ref{eomH1Vf}) while the Bianchi identities
for $H_{1,2}$ provide the second order field equations for the scalar fields $X, Y$.
The latter are obtained from the Lagrangian
\bea
{\cal L} &=& {R}\, {vol}_3 - 6\, d W \wedge {*} d W
-\tfrac32 e^{-4W}\,DY \wedge *DY
\nonumber\\[1ex]
&&{}
-18\, dV \wedge  * d V
- \ft12 e^{-12 V}
\left\{
dX\wedge *dX +8if dX\wedge *df^*-8if^* dX\wedge *df \right\}
\nonumber\\[1ex]
&&{}
- 8 e^{-6 V}\left(1+2e^{-6V} |f|^2 \right) df \wedge * df^*
+8 e^{-12 V}  \left\{ f^2 df^* \wedge * df^* + (f^*)^2 df \wedge * df  \right\}
\nonumber\\[1ex]
&&{}
 - 3 \alpha B_2 \wedge H_2 + (2 e^{-6 W } -  \tfrac{3}{2} \alpha^ 2 e^{-8 W})\, {vol}_{3}\;,
\label{L3D}
\eea
which thus provides an (on-shell) equivalent description of (\ref{diageinsteinL}).
In this form of the action, the vector field $B_1$ has disappeared while $B_2$ appears with a
Chern-Simons term such that its field equations yield the second of the duality equations (\ref{defH12}).
Inspection of the scalar kinetic terms shows that the geometry of the full scalar target space is given by
the K\"ahler manifold $SU(1,1)/U(1) \times SU(2,1)/U(2)$, whose two factors are parametrized by
$(W, Y)$ and $(V, f, f^*, X)$, respectively.
For completeness, we give an explicit construction of the latter space in appendix~\ref{app:coset} and
reproduce the kinetic term of (\ref{L3D}) in (\ref{coset}).

The K\"ahler structure of the scalar target space can be made explicit by introducing
the K\"ahler potential
\bea
K&=&  K_1(u,\bar{u},f,\bar{f})+ K_2(z,\bar{z})  ~\equiv~
-{\rm log} (\Re u-4 |f|^2) - 3\, {\rm log} (\Re z)\,,
\eea
in terms of the complex coordinates $\phi^\imath = \{u, f, z\}$ given by
\bea
u \equiv e^{6V} + 4\,|f|^2 - i X
\;,\qquad
z=e^{2W} + iY
\;,
\eea
in terms of which the target space metric is given by $g_{\imath\bar\jmath}=\partial_\imath\partial_{\bar\jmath} K$.
For later use (in particular for the coupling to fermions), it turns out to be useful
to define the complex dreibein $E_\imath{}^a$ according to
\bea
g_{\imath\bar\jmath} &=& E_\imath{}^a E_{\bar\jmath\,a}
\;,
\eea
(with $E_{\bar\imath\,a} = (E_\imath{}^a)^*$)
explicitly given as
\bea
E_\imath{}^a &=&
\left(
\begin{array}{ccc}
\ft12 e^{-6V} &0&0\\
-4 e^{-6V} \bar{f} & 2 e^{-3V}&0\\
0&0& \ft{\sqrt{3}}2 e^{-2W}
\end{array}
\right)
\;.
\label{dreibein}
\eea
It allows to express the scalar kinetic term of (\ref{L3D}) as
\bea
{\cal L}_{\rm kinetic} &=& -2 g_{\imath \bar\jmath} D_\mu \phi^\imath D^\mu \phi^{\bar\jmath}
~=~ -2 P_\mu^a P^\mu_a
\;,
\eea
with
\bea
P_\mu^a&=&E_\imath{}^a D_\mu \phi^\imath
\label{defP}\\[1ex]
&=&
\left\{
3 \partial_\mu V- \ft{i}2  e^{-6V}\left( \partial_\mu X+8 \Im(\bar{f}\partial_\mu f)   \right)\,,\;
2 e^{-3V } \partial_\mu f\,,\;
{\sqrt{3}}
\left(\partial_\mu W + \ft{i}2 e^{-2W} D_\mu Y \right)
\right\}
\;.
\nonumber
\eea

\subsection{Gauging and scalar potential}

In the general framework of~\cite{deWit:2003ja}, the couplings in the last line of (\ref{L3D}) can be understood
as a deformation of the unique ungauged $N=2$ theory with scalar target space $SU(1,1)/U(1) \times SU(2,1)/U(2)$.
In particular, the $\alpha$ dependent terms correspond to the gauging of the shift symmetry $Y\rightarrow Y+c$
according to the covariant derivative introduced after (\ref{defH12}). The remaining term in the scalar potential
descends from a holomorphic superpotential which is compatible with the $N=2$ structure. Identifying these
deformations from the bosonic Lagrangian then allows to reconstruct the full fermionic sector of the theory.

Explicitly, the new couplings due to the gauging\footnote{
In the notation of~\cite{deWit:2003ja}, this gauging is described by an embedding tensor
of the form $\Theta_{00}=\frac23\alpha$ which defines the minimal couplings
$D_\mu Y \equiv \partial_\mu Y + A^0_\mu \Theta_{00}$ upon introducing the vector fields
$A^0_\mu \equiv -3B_\mu$. Furthermore, w.r.t.\ to the conventions of this paper,
the full Lagrangian is rescaled as ${\cal L}\rightarrow2{\cal L}$.
}
are parametrized in terms of the real scalar dependent tensor $T$
\bea
T &=& \frac43 \alpha\, {\cal P}^0 {\cal P}^0  ~=~ \frac34 \alpha\, e^{-4W}
\;,
\label{defT}
\eea
expressed in terms of the moment map ${\cal P}^0\equiv\frac34 e^{-2W}$ of the gauged shift isometry.
In particular, its contribution to the scalar potential is given by
\bea
{\cal V}_T &=&
-8\,T^2 + 8\,g^{\imath\bar{\jmath}}\,\partial_\imath T\,\partial_{\bar{\jmath}} T ~=~
\ft32 \alpha^2 e^{-8W}
\;,
\eea
in accordance with (\ref{L3D}). This gauged supergravity is unique up to couplings induced
by a holomorphic superpotential ${\cal W}={\cal W}(u,f)$,\footnote{
Note that the required invariance of the holomorphic superpotential under the gauged isometries $Y\rightarrow Y+c$
implies that it does not depend on the complex variable $z$.}
whose contribution to the scalar potential is given by
\bea
{\cal V}_{\cal W} &=&
- 8\,e^K \, |{\cal W}|^2 + 2g^{\imath\bar{\jmath}}\,e^K D_\imath {\cal W} D_{\bar{\jmath}} {\cal W}^*
\;.
\label{VW}
\eea
With the explicit form of the K\"ahler metric given above, it is straightforward to verify that
the particular holomorphic superpotential
\bea
{\cal W}&=&\frac12(1+u)~=~=\frac12(1+e^\rho + 4 |f|^2 - i X)
\;,
\label{defW}
\eea
precisely yields the correct negative contribution ${\cal V}_{\cal W}=-2e^{-6W}$ to the scalar potential.
We should stress however that there are different choices for ${\cal W}$ which give rise to the same
contribution (\ref{VW}), i.e.\ the underlying $N=2$ structure is unique only up the choice of ${\cal W}$
satisfying ${\cal V}_{\cal W}=-2e^{-6W}$.
This is to be expected, as the global $SU(2,1)$ symmetry of the scalar sector
(such as the transformations (\ref{su21}))
is a symmetry of the
full bosonic Lagrangian but not of the holomorphic superpotential.
The entire $N=2$ fermionic sector can finally be reconstructed in terms of $T$ and ${\cal W}$ using
the general formulas given in~\cite{deWit:2003ja}.
We note in particular, the fermionic supersymmetry variations
\bea
\delta_\epsilon \psi_\mu &=&
\nabla_\mu\epsilon
- \ft14 i \left( e^{-6V} \left(\partial_\mu X+8\,\Im (\bar{f}\partial_\mu f )  \right)
-3 e^{-2W} D_\mu Y\right) \epsilon
\nonumber\\[1ex]
&&{}+
\ft34\alpha e^{-4W}\gamma_\mu \epsilon
+\ft12e^{-3W-3V}\, (1+e^{6V}+4|f|^2-i X) \,  \gamma_\mu \epsilon^*
\;,\nonumber\\[2ex]
\delta_\epsilon \lambda^a &=&
\ft12 \gamma^\mu P_\mu^a  \epsilon +  A^a  \epsilon +  B^a  \epsilon^*
\;,
\label{KS}
\eea
with $P_\mu^a$ from (\ref{defP}) and the scalar dependent tensors $A^a$, $B^a$
defined as
\bea
A^a &=& e^{-4W} \left\{0,0,\ft12\sqrt{3} \alpha \right\}\;,\label{AB}\\
B^a &=&
e^{-3W}\left\{
\ft14 e^{-3V}(1-e^{6V}+4|f|^2-i X),
- \bar{f},
\ft14 \sqrt{3} e^{-3V}(1+e^{6V}+4|f|^2-i X)
\right\}
\;.
\nonumber
\eea
These tensors likewise appear in the description of the Yukawa couplings of the fermionic fields.
W.r.t.\ \cite{deWit:2003ja} we have redefined the (complex)
spin 1/2 fermions as $\lambda^a \equiv E_\imath{}^a\chi^\imath$
with the dreibein from (\ref{dreibein})
and introduced the complex gravitino $\psi_\mu \equiv \psi_\mu^1+i \psi_\mu^2$\,.
Equations (\ref{KS}) define the Killing spinor equations of the the three-dimensional theory,
which we will employ in the following to identify various BPS solutions.

Let us finally note, that the scalar potential of the three-dimensional theory (\ref{L3D}) may be expressed
in terms of a {\em real} superpotential $F$ as follows
\bea
{\cal V} &=& - 8 \,F^2 +8\, g^{\imath\bar\jmath}\,\partial_\imath F\partial_{\bar\jmath}F
\;,
 \label{FV}
\eea
where we choose $F$ to be one of the eigenvalues of the gravitino mass matrix
\bea
F &=& -T \pm e^{K/2} |{\cal W}|
\;,\label{F1}
\eea
with $T$ and ${\cal W}$ from (\ref{defT}), (\ref{defW}). Remarkably, there is another choice for the
real superpotential
\bea
F &=& -\ft34 \alpha e^{-4W} \pm e^{-3W}
\;,
\label{F2}
\eea
which likewise generates the scalar potential via (\ref{FV}). Its existence may be related to some
fake supersymmetry structure of the theory.

\section{Equations of motion}
In this section we will begin by deriving the equations of motion in Einstein frame from the action (\ref{diageinsteinL}), before we exhibit some solutions which are the result of reducing known solutions in the literature. As expected, both the $AdS_3$ vacuum and the BTZ black hole \cite{BTZ} are solutions to the equations of motion. As we will later see, non-trivial rotations mean that it is not possible to embed the static charged BTZ \cite{chargedBTZ} black hole in $D=11$ supergravity using our ansatz.

The equations of motion derived from (\ref{diageinsteinL}) may be expressed as
\bea
0 &=& d (e^{12 V} * H_1) + 8 i df \wedge d {f}^*,  \nn
0 &=& d(e^{ 4 W} * H_2) +2 \alpha H_2,  \nn
0 &=& d(e^{-6 V} * df ) + i H_1 \wedge d f,  \nn
0 &=& d * d V + \tfrac{4}{3} e^{-6 V} df \wedge * df^* - \tfrac{1}{6} e^{12 V} H_1 \wedge * H_1,  \nn
0 &=& d * d W - \tfrac{1}{2} e^{W} H_2 \wedge * H_2
- e^{-6 W} vol_3 + \alpha^2 e^{-8 W} vol_3, \nn
R_{\mu \nu} &=& 6 \partial_{\mu} W \partial_{\nu} W +  18 \partial_{\mu} V \partial_{\nu} V  + 4 e^{-6V} (\partial_{\mu} f \partial_{\nu} f^{*} + \partial_{\mu} f^* \partial_{\nu} f) \nn &+& \eta_{\mu \nu} \left(\tfrac{3}{2} \alpha^2 e^{-8 W} - 2 e^{-6 W} \right) + \tfrac{1}{2} e^{12 V} ( H_{1 \mu \s } H_{1 \nu}^{~~\s} - \tfrac{1}{2} \eta_{\mu \nu} H_{1 \s_1 \s_2} H_{1}^{\s_1 \s_2})  \nn &+& \tfrac{3}{2} e^{4 W} ( H_{2 \mu \s } H_{2 \nu}^{~~\s} - \tfrac{1}{2} \eta_{\mu \nu} H_{2 \s_1 \s_2} H_{2}^{\s_1 \s_2}). \label{eom_einstein}
\eea

In the next subsection, we reduce a general class of known supersymmetric solutions from $D=11$ to $D=3$. These solutions also act as another rudimentary consistency check on some of the equations of motion.

\subsection{Supersymmetric solutions}
In addition to the supersymmetric $AdS_3$ vacuum, we can also check the Einstein frame equations of motion (\ref{eom_einstein}) by following the reduction of the intersecting M5-brane geometry of \cite{braneBH}. The $D=3$ solution is given by
\bea
ds^{2} &=& H^{3} r^{4} (du dv + K d u^2) + H^{6} r^{4} dr^2, \nn
H &=& 1+ \tfrac{\alpha}{r}, \quad K = 1 + \tfrac{Q}{r}, \quad {W} = \ln( r + \alpha),
\label{bh1}
\eea
where $ u = x -t, v = 2 t$ and apart from the metric, one has a scalar $W$ corresponding to the breathing mode of the $S^2$ (note $V=0$).
Indeed it is straightforward to check that (\ref{bh1}) provides a solution to the Killing spinor equations (\ref{KS}) with
$\epsilon=\frac{ir^{1/2}(\alpha+r)^{3/4}}{(Q+r)^{1/4}} \epsilon_0$ and a real constant spinor $\epsilon_0=\gamma^r \epsilon_0$.

We observe that this solution in $D=3$ has the same form as one of the black string solutions of \cite{d3bstring}. To see this, one can send $r \rightarrow -r$ in (\ref{bh1}) and then with $\alpha > 0$, the asymptotic geometry at $r=0$ is $AdS_3$, whereas the metric encounters a singularity at $r= \alpha$. Explicitly, the three-dimensional curvature scalar is of the form $R=\frac32\alpha(8r-\alpha)/(r-\alpha)^8$\,.

Here the constant $\alpha$ in the reduction (\ref{red_ansatz}) corresponds to the D4-charge, while $Q$ corresponds to the D0-charge that results when the geometry of \cite{braneBH} is reduced to type IIA.
Then taking a near-horizon decoupling limit similar to \cite{strexcl}, i.e. $r \rightarrow 0$ while $Q/r$ is kept constant, one recovers the usual rotating BTZ form
\be
ds^{2} = - \frac{(\rho^2 - \rho^2_*)^2}{\rho^2} dt^2 + \frac{4 \alpha^6 \rho^2}{(\rho^2 - \rho^2_*)^2} d \rho^2 + \rho^2( d x - \tfrac{\rho_*^2}{\rho^2} dt)^2 ,
\ee
where we have redefined \be \alpha^3 r = \rho^2 - \rho_*^2,  \quad \rho_*^2 = \alpha^3 Q. \ee
It is worth noting here the location of the horizon. Indeed, $\rho_{*}$ is given in terms of the product of three identical D4-charges $\alpha$ and the D0-charge $Q$. In the special case when $Q=0$, one recovers $AdS_3$ in the near-horizon limit with radius $\ell = 2 \alpha^3$.

A sterner check for the equations of motion can be obtained by working with the large class of $D=5$ supersymmetric black rings, black holes and supertubes compactified on $T^6$ from $D=11$ \cite{3chargeBH}. Our ansatz (\ref{red_ansatz}) means that when identifying $CY_3$ with $T^6 \equiv (T^2)^3$, each $T^2$ has to come with the same warp factor, i.e. $e^{2 V}$. The ansatz we have chosen for the reduction also confines us to direct products of $AdS_3$ with $S^2$ in $D=5$, though we remark that fibred solutions have appeared in \cite{deboer1}.

In contrast to \cite{deboer1}, we also confine ourselves to single-centred solutions and set D6-charge $p$ and rotation parameter $\omega$ to zero from the offset, so that the $D=5$ solution becomes a direct product of $AdS_3$ and $S^2$, thus enabling the two-sphere to be decoupled to give a $D=3$ solution. By rescaling correctly in going to Einstein frame one finds the following class of solutions
\bea
\label{susysoln}
ds^{2} &=& r^{4} \left[ - (dt + \mu dx)^2 + Z^3 ( dx^2 + dr^2) \right], \nn
H_2 &=& -d Z^{-1} \wedge dt - d \left[ Z^{-1} (\tfrac{1}{2} KL + M) \right] \wedge dz, \nn
W &=& \tfrac{1}{2} \ln (Z r^2),
\eea
where
\be
M = m_0 + \frac{m_1}{r}, \quad L = l_0 + \frac{l_1}{r}, \quad K = k_0 + \frac{k_1}{r},
\ee
are general harmonic functions with $Z$ and $\mu$ expressible in terms of them:
\bea
Z &=& K^2 + L, \nn
\mu &=& K^3 + \tfrac{3}{2}K L +M.
\eea
One final constraint comes from the setting the rotation parameter $\omega$ to zero:
\be
\label{constraint}
m_1 = \frac{3}{2} (k_1 l_0 - k_0 l_1 ).
\ee
The equations of motion (\ref{eom_einstein}) may be shown to be satisfied by making use of the relationship among the charges above (\ref{constraint}) and again setting the D4-charge, $k_1 = \pm \alpha$, providing us with a valuable consistency check on the $D=3$ Lagrangian.
For $\alpha=-1$ this solution is a solution to the three-dimensional Killing spinor equations (\ref{KS}) corresponding to a Killing spinor
$\epsilon = r \epsilon_0$ with constant $\epsilon_0$ satisfying $\epsilon_0^*=-\gamma^r \epsilon_0$, $\gamma_t \epsilon_0 = i \epsilon_0$\,.

Solutions constructed in this fashion are usually prone to causal pathologies, however this solution will be free of CTCs provided the $g_{xx}$ component of the metric has the correct signature. This can be guaranteed by ensuring that the following inequality holds everywhere
\be
Z^3 - \mu^2 \geq 0.
\ee

\subsection{Null-Warped $AdS_3$}
It is known that null-warped $AdS_3$ spacetimes with dynamical exponent $z$ exist as solutions \cite{strominger} to topologically massive gravity (TMG) \cite{TMG}\footnote{These spacetimes originally featured in \cite{Israel} and an overlap with geometries dual to non-relativistic CFTs also exists \cite{NRdual}.}. However, when the coupling to the Cotton tensor in TMG is switched off, these solutions may still be supported in the presence of a Maxwell Chern-Simons term \cite{sinha1}. In $D=3$ the solution to the Einstein equations may be written
\bea
ds^2_3 &=& - r^z dt^2 \pm 2 \beta r dt d x + \frac{\ell^2}{4} \frac{dr^2}{r^2}, \nn
A_{t} &=& \frac{2}{\alpha^2 \sqrt{3} z} \sqrt{r^z z (z-1)},
\eea
where the constant $\beta$ drops out of the equations of motion i.e. one can always rescale $x$, so $\beta$ is undetermined. The dynamical exponent $z$ is then set by the flux equations of motion and, for this particular case $z=4$.
Note now that this is the same as (\ref{susysoln}) when $\beta =1$ and only $k_1 = \pm \alpha$ is non-zero, so null-warped $AdS_3$ is a particular solution in this class and is supersymmetric.

Interestingly the dynamical exponent is the same as the non-relativistic solution identified in \cite{KE6} by performing the consistent truncation from $D=11$ on $KE_6$. However, as we have replaced $KE_6$ with $CY_3$, the final form of the uplifted solution is very different
\bea
ds^2_{11} &=& \frac{1}{\alpha^4} ds^2_{3} + \alpha^2 ds^2(S^2) + ds^2(CY_3), \nn
F^{(4)} &=&  \left(\alpha vol(S^2) + \frac{2 r}{\alpha^2}  dr \wedge dt  \right) \wedge J.
\eea

\subsection{A supersymmetric domain wall solution}
\label{sec:dw}

The explicit form of the Killing spinor equations (\ref{KS}) and the associated real superpotential (\ref{F1}) underlying the scalar potential according to (\ref{FV}) further allow to construct explicit domain wall  solutions within the three-dimensional theory. With the standard ansatz
\bea
ds^2 = e^{2A(r)}(-dt^2+dz^2)+dr^2
\;,
\eea
for the three-dimensional metric, with both vector fields set to zero, the Killing spinor equations reduce to
the differential equations
\bea
2e^{3V}(\alpha+e^{4W} W') &=& e^W+e^{6V+W}
\;,\qquad
{\rm sinh}(3V)~=~-3e^{3W}V'
\;,\nonumber\\[1ex]
A'&=& 2 e^{-3W}{\rm cosh}(3V)-\ft32\alpha e^{-4W}
\;.
\label{diff1}
\eea
This is a particular case of the standard first order equations
$(\phi^i)' = g^{ij}(\phi) \partial_j F$
for a theory with a scalar potential generated by a real superpotential $F$ 
via (\ref{FV}), see e.g.~\cite{Skenderis:1999mm}.
Equations (\ref{diff1}) can be explicitly integrated to give
\bea
e^W &=& \frac{3\alpha V}{{\rm sinh}(3V)}\;,\qquad
e^{2A} ~=~ \frac{216\,\alpha^3 V^3}{{\rm sinh}^4(3V)}
\;,
\eea
in terms of the single function $V$ which is defined by the first order equation
\bea
V' &=& -\frac{{\rm sinh}^4(3V)}{81\alpha^3 V^3}
\;.
\eea
In turn, this can be implicitly expressed in terms of polylogarithmic functions.\footnote{
More precisely, $V$ is implicitly defined by the equation
\bea
\alpha^{-3} r &=& 
\frac{18V^2 e^{6V} \left(1+4\, {\rm ln}(1-e^{-6V})\,{\rm sinh}^2(3V)\right)}{(1-e^{6V})^2}
+\frac{36 V^3 (1-3e^{6V})}{(1-e^{6V})^3}
\nonumber\\
&&{}
-{\rm ln}(1-e^{-6V})-{\rm Li}_3\, e^{-6V} -6V\left({\rm Li}_2\, e^{-6V}+\frac1{1-e^{6V}}\right)
\;.
\nonumber
\eea
}
For $r\rightarrow\infty$, this solution approaches the $AdS_3$ geometry
\bea
W \rightarrow {\rm log}\,\alpha\;, \quad V \rightarrow e^{-r \alpha^{-3}}\;,
\quad
A \rightarrow \ft12 \alpha^{-3}\,r 
\;,
\eea
while at $r=0$ the scalar fields and the space-time curvature diverge. It is possible to show that the equations of motion are also satisfied using the relationships above to write all equations in terms of $V$. 

\subsection{Numerical Black Holes}
\label{sec:nbh}

As the reader may appreciate from the proceeding sections, a non-zero gauge field $B_2$ leads to a rotating system, so henceforth we focus on solutions where this gauge field does not appear. We will also set $f=0$, but it should be noted that though the action bears some resemblance to an Einstein-Yang-Mills action, the static \textit{charged} BTZ black hole is not a solution. Indeed, one sees from (\ref{eom_einstein}) that one needs the presence of a non-constant scalar $V$, so that $H_1$ can be supported.

We next introduce the following black hole ansatz
\be
ds^2 = \frac{1}{z^2} \left[ - e^{2A} g dt^2 +  \frac{1}{g} dz^2 + d \varphi^2 \right],
\ee
where $A, g $ are functions of the radial direction $z$ only. Naturally the warp factors $W, V$ will also depend on $z$ and we will take the gauge fields $B_1$ to be purely electric
\be
B_1 = \phi(z) dt.
\ee

With this choice of ansatz the original equations of motion involving the matter content become respectively
\bea
 0 &=& \left[ e^{12 V -A} z \phi'\right]' , \nn
 0 &=& \left[\frac{g e^{A}  V'}{z}  \right]' - \tfrac{1}{6} e^{12 V-A} z (\phi')^2, \nn
\label{mattereom} 0 &=& \left[ \frac{g e^{A} W'}{z}  \right]'
+ \frac{e^{A}}{z^3} (e^{-6 W} - \alpha^2 e^{-8 W}).
\eea
The non-zero components of the Ricci tensor in orthonormal frame become
\bea
R_{tt} &=& e^{-A} z^2 \left[ \sqrt{g} z [e^{A} \frac{\sqrt{g}}{z} ]' \right]' - \sqrt{g} e^{-A} z^2 [e^{A} \frac{\sqrt{g}}{z}]', \nn
R_{zz} &=& - e^{-A} z^2 \left[ \sqrt{g} z [e^{A} \frac{\sqrt{g}}{z}]' \right]' + \tfrac{1}{2} z^2 \left[\frac{g}{z^2}\right]', \nn
R_{\varphi \varphi} &=&  \sqrt{g} z^2 e^{-A} [ e^{A} \frac{\sqrt{g}}{z}]' + \tfrac{1}{2} \left[\frac{g}{z^2} \right]' z^3,
\eea
meaning that the Einstein equations may then be expressed as
\bea
R_{tt} &=& -(\tfrac{3}{2} \alpha^2 e^{-8 W} - 2 e^{-6 W} )  \nn
R_{zz} &=& (\tfrac{3}{2} \alpha^2 e^{-8 W} - 2 e^{-6 W} ) + g z^2 \left( 6 W'^2 + 18 V'^2  \right), \nn
R_{\varphi \varphi} &=& (\tfrac{3}{2} \alpha^2 e^{-8 W} - 2 e^{-6 W} ) + \tfrac{1}{2} e^{12 V-2 A} z^4 (\phi')^2.
\eea

By combining $E_{tt}$ and $E_{zz}$ one gets the additional equation
\be
\label{addeq} {A'} = - 6 z \left[ (W')^2 + 3 (V')^2 \right].
\ee
As in \cite{sean2}, it is possible to show all equations are satisfied if the field  equations of motion (\ref{mattereom}), $E_{\varphi \varphi} = 0$ and (\ref{addeq}) are satisfied. This may be done by differentiating $E_{\varphi \varphi} = 0$ and showing that $E_{tt} = E_{zz} = 0$ may be obtained by using the other equations of motion. Therefore, one may ignore the equations involving second derivatives in $g$. We remark that if one sets $W=0$, then there is no solution to these equations, which may be verified by manipulating the equations.

Following \cite{sean2,cmjerome2}, one can solve the full set of equations numerically by integrating out from the horizon to the $AdS_3$ boundary. In doing so, we assume there is a horizon at $z=z_+$ defined by $g(z_+) = 0$ and consider a series solution:
\bea
A &=& a_0 + a_1 (z-z_+) + \cdots, \nn
g &=& g_{1} (z-z_+) + g_2 (z-z_+)^2 + \cdots, \nn
\phi &=& \phi_1 (z-z_+) + \phi_2 (z-z_+)^2 + \cdots, \nn
W &=& w_0 +  w_{1} (z-z_+) + \cdots,\nn
V &=&  v_0 +  v_{1} (z-z_+) + \cdots.
\eea
Ensuring that the equations of motion are satisfied at the horizon leads to expressions for the higher order terms in terms of the following set of parameters:
\be
z_+, \quad a_0, \quad \phi_1, \quad w_0, \quad v_0.
\ee

The temperature of the black hole is given by
\bea
T &=& - e^{- a_{b}} \frac{1}{4 \pi} \left[ e^{A} g' \right]_{z=z_+}, \nn
&=& e^{a_0- a_{b}} \frac{1}{ 8 \pi z_+} [4 e^{-6 w_0} -3 \alpha^2 e^{-8 w_0} - z_+^4 e^{12 v_0-2 a_0} \phi_1^2 ],
\eea
where we have used the subscript $b$ to denote the value at the boundary $z=0$.

From the numerics, we see that there are black hole solutions where $W$ starts off away from the $AdS_3$ minimum and falls down the potential to the bottom where it oscillates before settling, leading to an emergent conformal symmetry at the boundary $z=0$. The other breathing mode $V$ also saturates close to the boundary as is evident from Figure~1. Similar behaviour is also noted for $A$ near the boundary where it also approaches a constant. At the boundary the final form of the numeric solution is reminiscent of the charged BTZ black hole. In other words, the asymptotic form is
\bea
\phi &\sim& Q \ln z, \nn
g &\sim& \frac{1}{\ell^2} -M z^2 + e^{12 v_{b}-2 a_{b}} \frac{Q^2}{2} z^2 \ln z,
\eea
where $M$ is an integration constant related to the mass of the black hole.

\begin{figure}[ht]
  \label{conf}
  \centering
    \begin{tabular}{ll}
    \includegraphics[width=75mm]{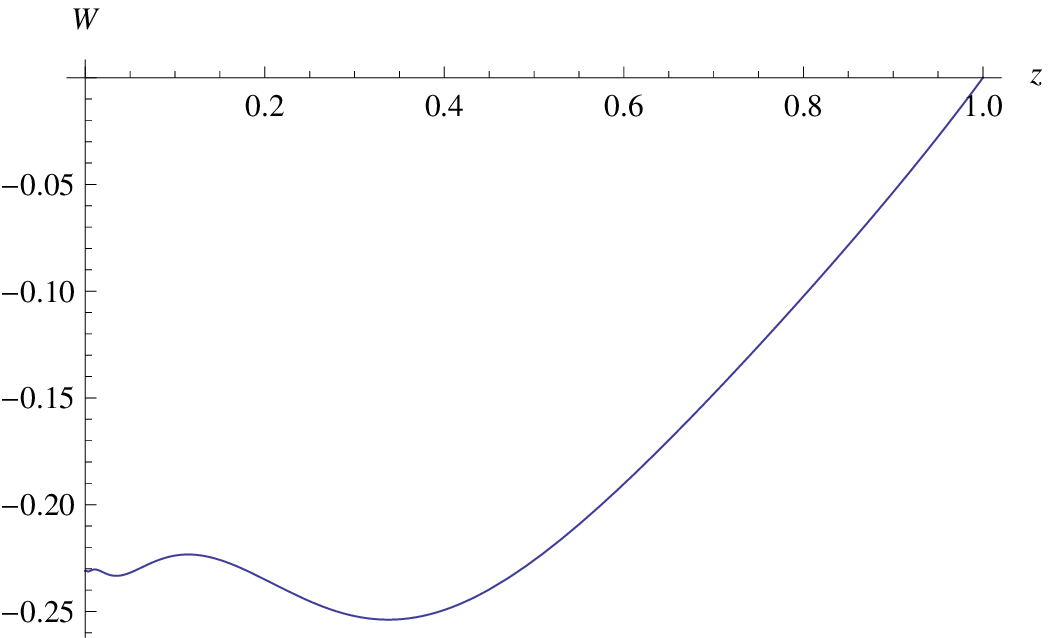} &\includegraphics[width=75mm]{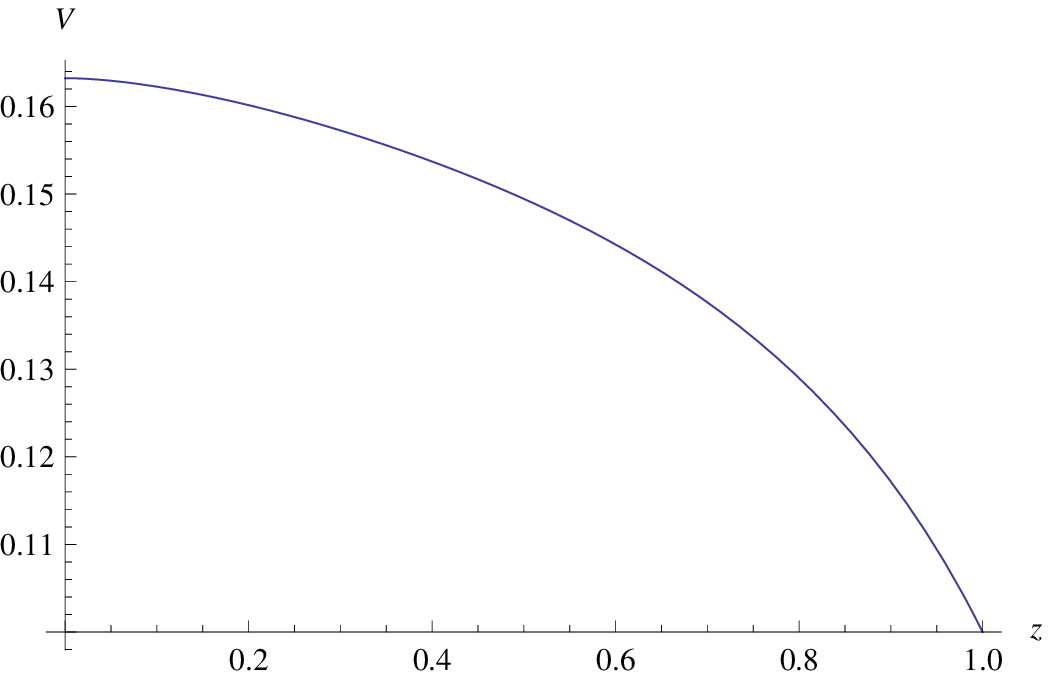}
     \end{tabular}
   \caption{The left-hand figure captures the emerging conformal symmetry as $W$ approaches its $AdS_3$ value at the boundary from a set initial starting value of $w_0 = 0$ at horizon ($AdS_3$ radius $\ell =1$ i.e. $W_{AdS} = -\tfrac{1}{3} \ln 2$). From the right-hand figure we see clearly that $V$ approaches a constant at the boundary. }
\end{figure}

We observe that the solution has scaling symmetries
\bea
e^{A} \rightarrow \beta^{-1} e^{A}, \quad t \rightarrow \beta t, \quad \phi \rightarrow \beta^{-1} \phi, \nn
z \rightarrow \beta^{-1} z,\quad (t,\varphi) \rightarrow \beta^{-1} (t, \varphi), \quad \phi \rightarrow \beta \phi.
\eea
Note that the first one may be chosen so that $A|_{z = 0} = a_{b} = 0$. We will henceforth work on the assumption that we will use this scaling symmetry to set $a_{b} =0$.

The Euclidean action $I_{E}$ for this hairy black hole may be determined in much the same way to \cite{sean2}. From the symmetries of the solution, the Einstein's equation implies that the Einstein tensor $G_{\varphi \varphi} \equiv R_{\varphi \varphi} - \tfrac{1}{2} g_{\varphi \varphi} R$ satisfies
\be
G_{\varphi \varphi} = \frac{1}{2 z^2} (\mathcal{L}-R).
\ee
Then using $R = - 2 G^{\mu}_{~\mu}$, one can write the Lorentzian action $S_0$ as
\be
S_0 = - \int d^3 x \sqrt{-g} \mathcal{L} =  2 \int d t d \varphi \int_{z=z_+}^{z=0} dz  \left[\frac{g e^{A}}{z^2} \right]' .
\ee
As $g(z_+) = 0$, the surface term at the horizon vanishes and we just get the surface term at $z=0$. The action thus diverges as $z \rightarrow 0$ and may be regulated in the standard fashion by incorporating a Gibbons-Hawking term and a counterterm to remove the $\ln z$ divergence term \cite{finiteS,finiteSgauge}
\be
S_{1} = \int_{z \rightarrow 0} dt d \varphi \sqrt{-\gamma} \left(- K + \frac{e^{12 v_{\infty}}}{2} n_{\mu} H_1^{\mu \nu} B_{1\nu} \right),
\ee
where $\gamma$ is the induced boundary metric at $z= 0$, $n_{\mu}$ corresponds to the outward pointing unit normal vector to the boundary and $K = \g^{\mu \nu} \nabla_{\mu} n_{\nu}$ denotes the extrinsic curvature. The presence of the second term can be motivated by looking at the variation of the Yang-Mills term $H_{1\mu \nu} H_{1}^{\mu \nu}$ at the boundary. Finally, the Euclidean action $I_{E}$ may be obtained from the total finite action $S_{t} \equiv S_0 + S_1$ by analytic continuation through the redefinitions $ t = - i \tau$, $i S_{t} = -I_{E}$.

\subsection{Conductivity}
Linear response theory can be used to determine the various conductivities of the dual field theories. One proceeds \cite{sean1, sean2} by considering small perturbations of the form
\bea
\delta g_{t \varphi} &=& h_{t \varphi}(z) e^{-i \omega t}, \nn
\delta B_1 &=& b(z) e^{-i \omega t} d \varphi,
\eea
with all other fields retaining their background values. Linearising the equations of motion, one finds two independent equations
\bea
b'' + b' \left( 12 V' + A'+ \frac{g'}{g} + \frac{1}{z} \right) + b  e^{-2 A}  \left(\frac{\omega^2}{g^2} - e^{12V} \frac{z^2 \phi'^2}{g}  \right) &=& 0  , \nn
h_{t \varphi}' + \frac{2}{z} h_{t \varphi} + e^{12 V} \phi' b &=& 0.
\eea
Remarkably the Einstein equations are first order with the second order Einstein equations being implied by the equations of motion.

The equation for the gauge fluctuation may now be integrated numerically in the black hole background described earlier. In proceeding, one typically introduces infalling boundary conditions at the horizon through the ansatz \be b(z) = g^{-i \omega/(e^{a_0} g_1)} \rho(z) = g^{-i \omega/(4 \pi T)} \rho(z).  \ee
This takes care of oscillations at the horizon and leaves one with a differential equation to be solved for $\rho(z)$. As described in \cite{sean1}, one then proceeds to expand $\rho(z)$ at the horizon in a Taylor series to ensure that the equation is satisfied there, before integrating from the horizon out to the boundary.

\begin{figure}[ht]
  \centering
    \begin{tabular}{ll}
    \includegraphics[width=75mm]{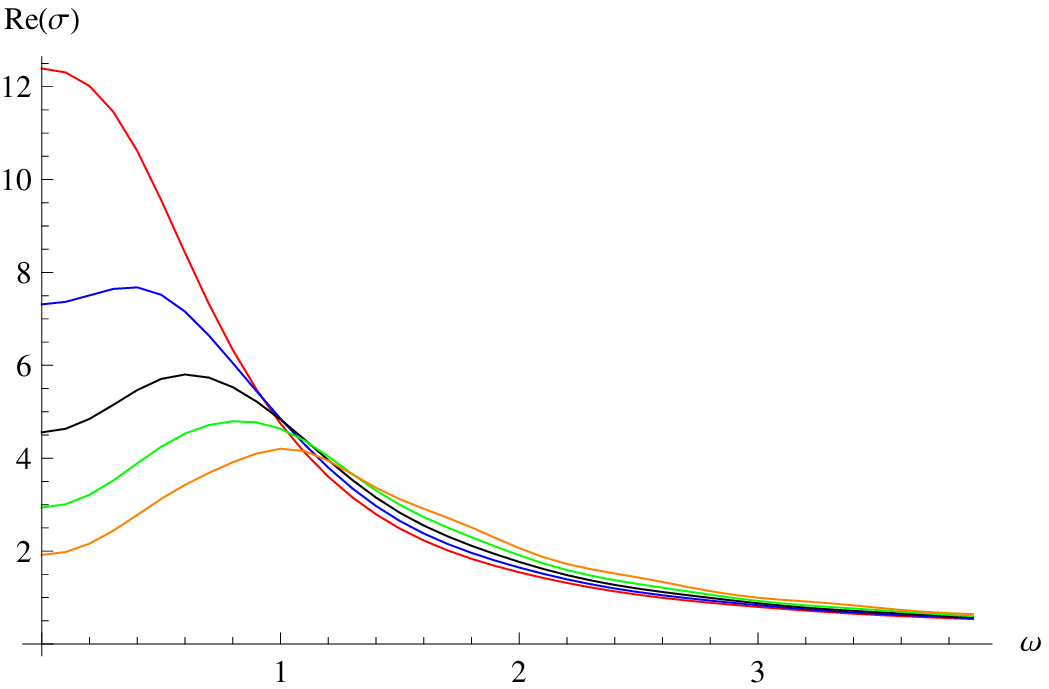} &\includegraphics[width=75mm]{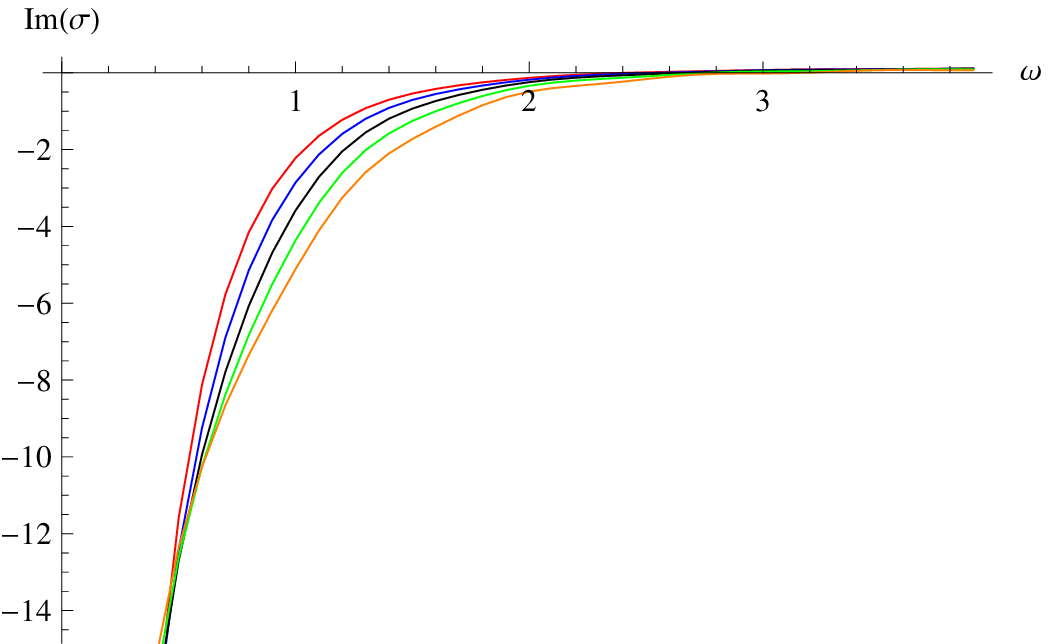}
     \end{tabular}
   \caption{The left and right graphs here illustrate the results of calculating  Re($\sigma$) and Im($\sigma$) at temperatures of $T=0.042$ (Red), $T=0.038$ (Blue), $T=0.033$ (Black), $T=0.028$ (Green) and $T=0.022$ (Orange).}
\end{figure}

In order to determine the conductivity, one simply has to note the form of the solution near the boundary. Asymptotically, $\rho$ is of the form $\rho \sim \rho_0 + \rho_1 \ln z$, where the complex valued constants can easily be read off from the numerical solution. As explained in some detail in \cite{Ren} (see also \cite{Hung,Maity}), the Green's function may then be defined by the ratio $G = - \rho_0/\rho_1$. It is also worth observing that the Green's function has an ambiguity due to the logarithmic term and may be shifted by a constant $\ln C$. The conductivity $\sigma(\omega)$ may then be determined from
\be
\sigma(\omega) = \frac{i}{\omega} \frac{\rho_0}{\rho_1}.
\ee

Since the conductivity is calculated at the boundary and the numerical solution asymptotes to the charged BTZ solution there, it is not surprising to find considerable agreement with the results of \cite{Maity, Ren}. Figure 2 shows the result of a calculation of the conductivity at various temperatures.

\section{Discussion}
In this work, motivated primarily by AdS/CMT considerations, we have exhibited one of the first examples of a consistent reduction from $D=11$ supergravity to $D =3$ supergravity at the level of the bosonic equations of motion. The choice of internal geometry, $S^2 \times CY_3$, was inspired by the observation in \cite{eoin1} that the MSW geometry should be the only example in its $SU(3)$-structure class, and also to some extent by the simplicity of the undeformed wrapped M5-brane geometry. Though one may have naively expected multiple $AdS_3$ vacua from the offset, the reduction presented here leads to a simple scalar potential with a single supersymmetric $AdS_3$ vacuum.

The scalar potential proves instrumental in uncovering rich $N=2$ supersymmetric structure up to the determination of a holomorphic superpotential term. As we have seen, once the supersymmetric structure is illuminated, the full Lagrangian may be easily read off from the established $D=3$ literature. Throughout this work, we have performed numerous consistency checks and have made use of supersymmetric solutions reduced from $D=11$ leading to black strings, warped $AdS_3$, among other solutions in $D=3$. These solutions have also served to check the fermionic extension by ensuring that the Killing spinor equations are satisfied in $D=3$.

In the final part of this paper, we have constructed an electrically charged hairy BTZ solution numerically in the reduced $D=3$ theory. As a first step in AdS/CMT applications, we have computed the conductivities, but the identification of this solution may be regarded as an important milestone in examining the physics of the dual strongly coupled CFT from the perspective of M-theory. Of particular interest in this regard will be applying the techniques of \cite{nonfermi} to study the spectral functions of the fermions in the hairy BTZ black hole. Note here that the fermion mass is motivated from the supersymmetric Lagrangian, instead of being plucked from thin air. Any observation of characteristic Luttinger liquid behaviour would offer support to the notion that applying AdS/CFT to condensed matter is not so far-fetched.

In other work, it would certainly be interesting to extend this study to find consistent truncations from either $D=10$ or $D=11$ supergravity to $D=3$ with multiple $AdS_3$ vacua permitting interpolating solutions dual to RG flows. Such solutions have been discussed in \cite{Berg:2001ty} in the context of $D=3$ gauged supergravity. Ideally, such flows should preserve some supersymmetry, as when supersymmetry is broken, doubts are raised over stability \cite{cmstability}.
A systematic analysis of $D=3$ supersymmetric solutions of this type may proceed along the lines of \cite{Deger:2010rb}.

\section*{Acknowledgements}
 We are grateful for early discussions with Oscar Varela and Patta Yogendran, and we would also like to thank Sean Hartnoll, Nakwoo Kim, Aninda Sinha, Mario Trigiante,  Matthias Wapler and Hossein Yavartanoo for their kind input at various stages. In addition we thank Oscar for comments on a final draft. E\'OC would like to also express gratitude to NUI Galway and ENS Lyon for warm hospitality at the later stages of this project. The work of H.S.\ is supported in part by the Agence Nationale de la Recherche (ANR).

\appendix

\section{Details of consistent truncation}
The Bianchi $d F^{(4)} = 0 $ is satisfied provided $\alpha$ and $\beta$ are constants and
\bea
d H_1 &=& - \gamma vol(M_3), \nn
H_2 &=& d B_2,
\eea
where $B_2$ is a one-form potential. The flux equations of motion
\be
d (* F^{(4)} ) + \tfrac{1}{2} F^{(4)} \wedge F^{(4)} = 0,
\ee
are then satisfied if
\bea
\beta &=& 0, \nn
\label{fluxeq1} d (e^{ -2 U + 6V} * H_1) + 8 i df \wedge d {f}^* &=& 0, \nn
d (e^{ 6 V} \gamma) &=& 0, \nn
\label{fluxeq2} d(e^{ 2 U + 2 V} * H_2) +2 \alpha H_2 &=& 0, \nn
d(e^{2U} * df ) + i H_1 \wedge d f &=& 0.
\eea
In deriving these expressions we have used
\bea
* \Omega &=& i \Omega, \quad
* \bar{\Omega} = -i \bar{\Omega}, \nn
\Omega \wedge \bar{\Omega} &=& - 8 i vol(CY_3).
\eea
The next step is to consider the Einstein equations
\be
R_{MN} = \tfrac{1}{12} F^{(4)}_{MP QR} F_{N}^{(4)~PQR} - \tfrac{1}{144} g_{MN} F^{(4)}_{PQRS} F^{(4)~PQRS}.
\ee
Using the elfbein
\be
e^{\mu} = \bar{e}^{\mu}, \quad e^{\alpha} = e^{U} \bar{e}^{\alpha},  \quad e^{i} = e^{V} \bar{e}^i,
\ee
where $\mu = 0,12$, $\alpha = 3,4$ and $i = 5,\cdots 10$, the non-zero components of the Ricci-tensor in orthonormal frame may be expressed as
\bea
R_{\mu \nu} &=& \bar{R}_{\mu \nu} - 2 (\nabla_{\nu} \nabla_{\mu}U + \partial_{\mu} U \partial_{\nu} U ) - 6 (\nabla_{\nu} \nabla_{\mu} V+ \partial_{\mu} V \partial_{\nu} V ), \nn
R_{\alpha \beta} &=& \delta_{\alpha \beta} \left[ e^{-2 U} -\nabla_{\rho} \nabla^{\rho}U - 2  \partial_{\rho} U \partial^{\rho} U - 6 \partial_{\rho} U \partial^{\rho} V\right], \nn
R_{ij} &=& \delta_{ij} \left[ - \nabla_{\rho} \nabla^{\rho} V - 6 \partial_{\rho} V \partial^{\rho} V - 2 \partial_{\rho} U \partial^{\rho} V \right].
\eea
We may now use the above Ricci tensors in writing out the Einstein equations. The field strength squared term may be expressed
\bea
\tfrac{1}{4!} F^{(4)}_{PQRS} F^{(4)~PQRS} &=& 3 \alpha^2 e^{-4 U - 4 V} + \tfrac{1}{2} e^{-4U} H_{1\rho \sigma} H^{\rho \sigma}_1 + \tfrac{3}{2} e^{-4V} H_{2 \rho \sigma} H_2^{\rho \sigma} \nn &+& 16 e^{-6 V} \partial_{\rho} f \partial^{\rho} f^*.
\eea
From the directions along the sphere we see that $\gamma = 0$ for consistency, with the final equation being
\bea
\label{eqn1}
\nabla_{\rho} \nabla^{\rho} U &+& 2 \partial_{\rho} U \partial^{\rho} U + 6 \partial_{\rho} U\partial^{\rho} V - e^{-2U} + \alpha^2 e^{-4 U -4 V} \nn &+& \tfrac{1}{6} e^{-4 U} H_{1\rho \sigma} H_1^{\rho \sigma} - \tfrac{1}{4} e^{-4 V} H_{2\rho \sigma} H_2^{\rho \sigma} - \tfrac{8}{3} e^{-6 V} \partial_{\rho} f \partial^{\rho} f^* = 0.
\eea
Then from the directions along the $CY_3$ we find the following expression
\bea
\label{eqn2}
\nabla_{\rho} \nabla^{\rho} V &+& 6 \partial_{\rho} V \partial^{\rho} V + 2 \partial_{\rho} U \partial^{\rho} V - \tfrac{1}{12} e^{-4 U} H_{1\rho \sigma} H_1^{\rho \sigma} + \tfrac{4}{3} e^{-6 V} \partial_{\rho} f \partial^{\rho} f^* = 0. \nn
\eea
Finally, along the three-dimensional space we have the 3d Einstein equation
\bea
\label{eqn3}
\bar{R}_{\mu \nu} &=& 2 ( \nabla_{\nu} \nabla_{\mu}U + \partial_{\mu} U \partial_{\nu} U) + 6 (\nabla_{\nu} \nabla_{\mu} V + \partial_{\mu} V \partial_{\nu} V ) - \tfrac{1}{2} \eta_{\mu \nu}  \alpha^2 e^{-4 U -4 V} \nn &+& \tfrac{1}{2} e^{-4 U} \left(H_{1 \mu \rho} H_{1\nu}^{~~\rho} - \tfrac{1}{6} \eta_{\mu \nu} H_{1 \rho \sigma} H_{1}^{\rho \sigma} \right)  + \tfrac{3}{2} e^{-4 V} \left(H_{2 \mu \rho} H_{2\nu}^{~~\rho} - \tfrac{1}{6} \eta_{\mu \nu} H_{2 \rho \sigma} H_{2}^{\rho \sigma} \right) \nn &+& 4 e^{-6 V} \left( \partial_{\mu} f \partial_{\nu} f^* + \partial_{\nu} f \partial_{\mu} f^* - \tfrac{2}{3} \eta_{\mu \nu} \partial_{\rho} f \partial^{\rho} f^* \right).
\eea
Observe that equation (\ref{eqn1}) and (\ref{eqn2}) may be repackaged as
\bea
\label{diffeq1}
d(e^{2 U + 6 V} * d U) &-& e^{6V} vol_3 + \alpha^2 e^{-2 U + 2 V} vol_3 - \tfrac{8}{3} e^{2U} df \wedge * df^* \nn &+& \tfrac{1}{3} e^{-2 U + 6 V} H_1 \wedge * H_1 - \tfrac{1}{2} e^{2 U+2 V} H_2 \wedge *H_2 = 0, \\
\label{diffeq2} d(e^{2 U + 6 V} * d V) &-& \tfrac{1}{6} e^{-2 U + 6 V} H_1 \wedge * H_1 + \tfrac{4}{3} e^{2U} df \wedge * df^* = 0.
\eea

\section{Coset space $SU(2,1)/U(2)$}
\label{app:coset}

In this appendix, we give an explicit construction of the scalar target space $SU(2,1)/U(2)$.
The group $SU(2,1)$ is defined as the set of matrices $U$ satisfying $U\eta\, U^\dagger=\eta$,
with $\eta={\rm diag}\{-1,1,1\}$. Its maximal compact subgroup is $U(2)=U(1)\times SU(2)$.
A basis of generators of the latter is given by matrices
\bea
T_{U(1)} \equiv
\left(
\begin{array}{cc}
-2i&\\
&i I_2
\end{array}
\right)
,\qquad
T_{SU(2),k} \equiv
\left(
\begin{array}{cc}
0&\\
&i \sigma_k
\end{array}
\right)
\;,
\label{compact}
\eea
with Pauli matrices $\sigma_k$ and the $2\times2$ identity matrix $I_2$\,.
The full algebra $\mathfrak{su}(2,1)$ is spanned by the compact generators (\ref{compact})
together with the four generators
\bea
T_0 \equiv
\footnotesize{\left(
\begin{array}{ccc}
0\!&\!1\!&\!0\\
1\!&\!0\!&\!0\\
0\!&\!0\!&\!0
\end{array}
\right)}
,\;
T_{+1} \equiv
\footnotesize{\left(
\begin{array}{ccc}
0\!&\!0\!&\!0\\
0\!&\!0\!&\!0\\
-i\!&\!i\!&\!0
\end{array}
\right)}
,\;
T'_{+1} \equiv
\footnotesize{\left(
\begin{array}{ccc}
0\!&\!0\!&\!i\\
0\!&\!0\!&\!i\\
0\!&\!0\!&\!0
\end{array}
\right)}
,\;
T_{+2} \equiv
\footnotesize{\left(
\begin{array}{ccc}
i\!&\!-i\!&\!0\\
i\!&\!-i\!&\!0\\
0\!&\!0\!&\!0
\end{array}
\right)}
\;.
\eea
The subscript refers to the grading defined by the adjoint action of the noncompact generator $T_0$.
We choose to parametrize a coset representative ${\cal V}$ in triangular gauge as
\bea
{\cal V} &=&
{\rm exp}(\ft12X T_{+2})\;
{\rm exp}(2f^* T_{+1} + 2f T'_{+1})\;
{\rm exp}(3V T_0)
\;.
\label{triangular}
\eea
According to the coset space structure, the left-invariant scalar current
may be decomposed into
\bea
J_\mu &\equiv&
{\cal V}^{-1} \partial_\mu {\cal V} ~\equiv~ Q_\mu + P_\mu
\;,
\eea
where $Q_\mu \equiv \frac12(J_\mu - J_\mu^\dagger)$ and
$P_\mu\equiv \frac12(J_\mu + J_\mu^\dagger)$ live in the
compact part of the algebra spanned by (\ref{compact}), and its orthogonal complement,
respectively.
The target space metric on the coset manifold $SU(2,1)/U(2)$ is given by
\bea
- {\rm Tr} (P^\mu P_\mu)
&=&
 -18\, \partial^\mu V \, \partial_\mu V -\ft12 e^{-12V}\, \partial^\mu X \partial_\mu X
-8e^{-\rho}(1+2 e^{-6V} ff^*)\, \partial^\mu f\, \partial_\mu f^*
\nonumber\\[1ex]
&&{}
+4 e^{-12V}
\left( i f^* \partial^\mu X \partial_\mu f+ 2 (f^*)^2 \partial^\mu f \partial_\mu f + c.c.\right)
\;,
\label{coset}
\eea
which precisely reproduces the corresponding kinetic term of the Lagrangian (\ref{L3D}).

The $SU(2,1)$ isometry group acts on the four coordinates in a non-linear way.
Its action may be made manifest in the triangular gauge (\ref{triangular}) by acting on the
coset representative by left multiplication
\bea
{\cal V} &\rightarrow& G \,{\cal V}\, H_G
\;,
\eea
with an $SU(2,1)$ element $G$, and a compensating right multiplication with $H_G\in U(2)$
that restores the triangular gauge. An example of such a non-trivial (Ehlers-type) transformation
is given by
\bea
e^{6V} &\rightarrow& \frac{e^{6V}}{1+8e^{6V} |\omega|^2+16 |\omega|^4 (e^{12V}+X^2 )}\;,
\nonumber\\[.5ex]
{f} &\rightarrow& \frac{\omega\, (e^{6V}-i X)}{1+4 |\omega|^2 (e^{6V}-iX )}\;,
\nonumber\\
{X} &\rightarrow& \frac{X}{1+8e^{6V} |\omega|^2+16 |\omega|^4 (e^{12V}+X^2 )}
\;,
\label{su21}
\eea
for complex constant $\omega$, where for simplicity we have only given the action
onto a solution with vanishing $f$.
By means of this transformation one may e.g.\ construct solutions with non-trivial $f$
from the solutions given in sections~\ref{sec:dw} and \ref{sec:nbh}.

\end{document}